\documentclass[%
 reprint,
superscriptaddress,
 amsmath,amssymb,
 aps,
]{revtex4-2}

\usepackage{graphicx}
\usepackage{dcolumn}
\usepackage{bm}
\usepackage{cleveref}

\usepackage{xcolor}

\begin{document}

\preprint{APS/123-QED}

\title{Mapping strain and structural heterogeneities around bubbles in amorphous ionically conductive Bi$_2$O$_3$}

\author{Ellis Rae Kennedy}
\thanks{These authors contributed equally to this work.}%
\affiliation{Materials Physics and Applications Division, Los Alamos National Laboratory, Los Alamos, NM, USA}%

\author{Stephanie M. Ribet}
\thanks{These authors contributed equally to this work.}%
\affiliation{National Center for Electron Microscopy, Molecular Foundry, Lawrence Berkeley National Laboratory, Berkeley, CA, USA}%

\author{Ian S. Winter}
\affiliation{Sandia National Laboratories, Livermore, CA, USA}%

\author{Caitlin A. Kohnert}
\affiliation{Materials Science and Technology Division, Los Alamos National Laboratory, Los Alamos, NM, USA}%

\author{Yongqiang Wang}
\affiliation{Materials Science and Technology Division, Los Alamos National Laboratory, Los Alamos, NM, USA}%

\author{Karen C. Bustillo}
\affiliation{National Center for Electron Microscopy, Molecular Foundry, Lawrence Berkeley National Laboratory, Berkeley, CA, USA}%

\author{Colin Ophus}
\affiliation{Department of Materials Science and Engineering, Stanford University, Palo Alto, CA, USA}%

\author{Benjamin K. Derby }
\affiliation{Materials Physics and Applications Division, Los Alamos National Laboratory, Los Alamos, NM, USA}%

\date{\today}

\begin{abstract}

While amorphous materials are often approximated to have a statistically homogeneous atomic structure, they frequently exhibit localized structural heterogeneity that challenges simplified models.
This study uses 4D scanning transmission electron microscopy to investigate the strain and structural modifications around gas bubbles in amorphous Bi$_2$O$_3$ induced by argon irradiation.
We present a method for determining strain fields surrounding bubbles that can be used to measure the internal pressure of the gas.
Compressive strain is observed around the cavities, with higher-order crystalline symmetries emerging near the cavity interfaces, suggesting paracrystalline ordering as a result of bubble coarsening.
This ordering, along with a compressive strain gradient, indicates that gas bubbles induce significant localized changes in atomic packing.
By analyzing strain fields with  maximum compressive strains of 3\%, we estimate a lower bound on the internal pressure of the bubbles at 2.5 GPa.
These findings provide insight into the complex structural behavior of amorphous materials under stress, particularly in systems with gas inclusions, and offer new methods for probing the local atomic structure in disordered materials.
Although considering structural heterogeneity in amorphous systems is non-trivial, these features have crucial impacts on material functionalities, such as mechanical strength, ionic conductivity, and electronic mobility.
\end{abstract}

\maketitle

\section{Introduction}

The properties of crystalline materials are determined by their unit cell and a set of symmetry operations.
This concise description of structure does not extend to amorphous materials, which lack long-range order, translational symmetry, and rotational symmetry~\cite{sausset2011characterizing,kuji2025clarification,straus2016self}.
As a result, assessing structural variations in amorphous materials has been a long standing challenge. 
The importance of understanding the structure of amorphous materials is underscored by their broad range of applications, including in advanced battery technologies~\cite{guo2023amorphous, ding2024amorphous}, high-performance coatings~\cite{bhushan1999chemical, hu2014amorphous}, and functional materials~\cite{croissant2020synthetic}. 
Developing methods to understand the structural attributes of novel amorphous materials is essential for linking their properties to functionality for further technological development.

There are a number of synthesis and treatment pathways leading to the amorphization of a material.
For example, irradiation has been shown to induce phase transformations and amorphization in a wide range of structures, including ceramics ~\cite{hobbs1994radiation,kennedy2024insights,jenczyk2022unexpected}, semiconductors~\cite{pelaz2004ion}, and metallic alloys~\cite{benyagoub1988amorphization}.
Other processes, such as shock loading in metals~\cite{li2022amorphization} and electrical biasing in perovskite structures, can result in full or partial amorphization~\cite{kim2021imaging}.
Ion irradiation is unique among these material treatments as the ability to fine-tune processing parameters makes it a controlled method to induce amorphization. 
The ionic conductor Bi$_2$O$_3$ is a widely used material in applications such as solid oxide fuel cells, catalysis, solid electrolytes, antibacterial coatings, and optoelectronics, largely due to its ionic conductivity and electronic properties~\cite{kennedy2024insights,mane2024review,zahid2021review,lei2020ligaos,fruth2007structural,carlsson2002theoretical}.
Bi$_2$O$_3$ takes on a number of polymorphs with different material properties, and recent efforts have explored ion-irradiation as a means to stabilize the functional high-temperature delta phase at room temperature. These studies instead produced an amorphous phase, which we explore here as a heterogenous amorphous system~\cite{wang2023intrinsic,kennedy2024insights}.
Despite this, amorphous Bi$_2$O$_3$ remains a valuable model system for studying the structural evolution of ionically conducting oxides, particularly in environments where radiation exposure may influence material performance.
In this study, the stable monoclinic $\alpha$-Bi$_2$O$_3$ is subjected to Ar$^+$ ion irradiation, serving as a representative amorphous system for structural analysis.

Due to the wide applicability of amorphous materials, such as the bismuth oxide addressed here, there is significant interest in characterizing complex amorphous structures.
Although the structural analysis of amorphous materials differs from that of crystalline materials, several analogous approaches exist.
For instance, two-body and multi-body distribution functions are commonly applied to determine the probability that two or more atoms are separated by a specific set of vectors.
Short-range order (SRO) is often analyzed through two-body statistical methods such as radial distribution functions.
These methods provide insights into the local atomic arrangements of amorphous materials, where long-range order is absent~\cite{sausset2011characterizing,kuji2025clarification}.

Nonetheless, one of the challenges in characterizing amorphous materials is that they are inherently heterogeneous in structure~\cite{im2018direct,kennedy2024exploring}.
Unfortunately many existing experimental methods rely on the assumption that the atomic packing in amorphous systems, though random, is statistically homogeneous over a given area or volume.
However, this assumption of ``randomness'' is not typically valid, as amorphous materials often exhibit localized structural heterogeneity, namely medium-range ordering and small-scale variations in atomic arrangement~\cite{gibson1997diminished}. 
These variations are difficult to capture using traditional models based solely on random packing, such as those based on rapid quenching or conventional packing density functions.
It is increasingly recognized that such models fail to fully represent the structural nuances inherent to highly disordered materials and limit understanding of how variations in the structure affect functional properties~\cite{gibson1997diminished}.

The challenge of understanding local fluctuations in the structure of a material is especially pronounced when considering strain. 
In crystalline materials, strain is defined as the deviation in atomic spacing from a reference lattice.
In amorphous materials, strain is typically quantified as deviations in the atomic spacing from a reference value.
Although amorphous strain measurements have been previously performed using X-ray~\cite{scudino2015structural, poulsen2005measuring} and electron techniques~\cite{ebner2016local, pekin2019direct, gammer2018local}, these analyses have generally focused on metallic glasses, with less emphasis on more complex systems such as oxides. In these materials, heterogeneity exists both between coexisting phases and within individual amorphous regions.

The small probe size and versatility concomitant to scanning transmission electron microscopy, makes it an ideal tool to study local fluctuations in amorphous solid oxide structures.
In this study, we present a methodology for mapping strain in nanoscale volumes of amorphous materials surrounding bubbles embedded in a solid matrix using 4D scanning transmission electron microscopy (4D-STEM)~\cite{ophus2019four}.
The diffraction patterns obtained from 4D-STEM provide information on the relative distances between atoms, enabling the mapping of strain around high-pressure Ar bubbles and voids formed by Ar escape.

We are able to estimate the maximum strain at the cavity edges and to map the strain fields around these cavities.
From these strain measurements, we approximate the internal pressure of the bubbles (filled cavities), despite their small radial size, which can complicate gas pressure estimations in solid systems.
Additionally, we demonstrate that the same diffraction patterns can be used to map variations in crystalline symmetries embedded within the amorphous matrix.

\begin{figure}[t]
\centering
\includegraphics[width = 0.45\textwidth]{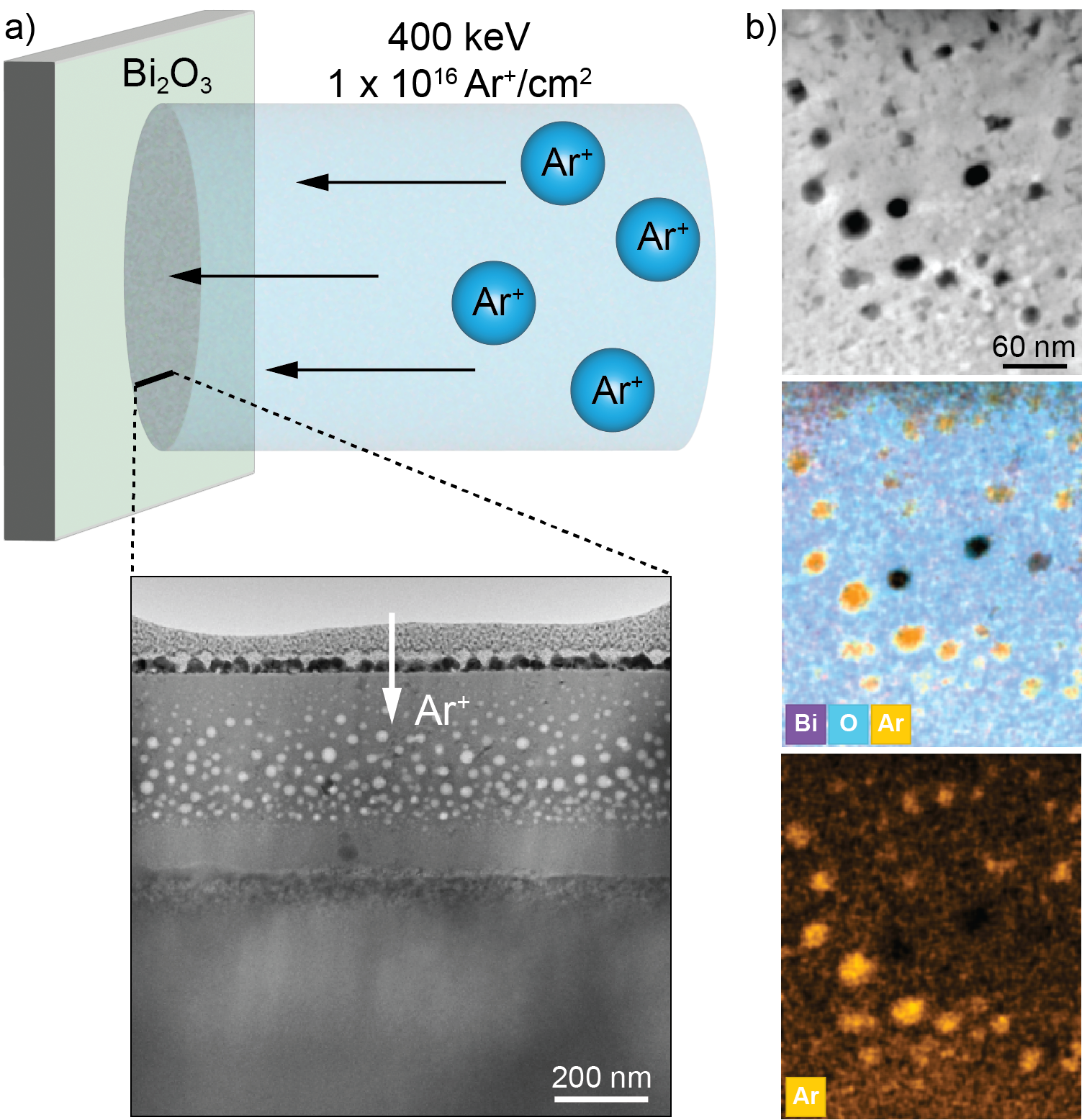}
  \caption{(a) Schematic depiction of 400 keV Ar$^+$ irradiation of $\alpha$-Bi$_2$O$_3$. The surface of the material was uniformly irradiated within the path of ion beam. The resulting damage profile includes a 440 nm amorphized layer with dislocation loops embedded in the parent crystal at the end of the damage range. Ar gas bubbles and voids of various sizes are present in the amorphous layer. These bubbles provide interfaces resulting in local variations in the amorphous structure of the material. (b) EDS from the top 25 to 315 nm of the amorphized region shows that some cavities are (unfilled) voids, while the majority are (filled) bubbles. The voids likely result from gas escaping during FIB sample preparation.}
  \label{fig:overview}
\end{figure}

Under 400 keV Ar$^+$ irradiation, a fully amorphous layer of Bi$_2$O$_3$ forms, with embedded Ar bubbles, (Figure~\ref{fig:overview}a-b) offering a system with well-defined heterostructures for investigation. Furthermore, defining equations of state for bubbles within solids and measuring the internal pressure of these bubbles has been a topic of interest for radiation effects research.

This work not only offers a novel approach for mapping strain in solid oxide materials but also underscores the need to consider amorphous materials as heterogeneous systems.
Structural deviations, such as strain and short- and medium-range ordering that we have uncovered, can have significant implications for material properties~\cite{tuller1989amorphous,zhu2024uncovering}.
In solid oxides like Bi$_2$O$_3$, variations in structure can influence ionic mobility, bulk conductivity, and optoelectronic behavior, making the understanding of these structural features crucial for optimizing their performance across applications~\cite{chen2019effect}.

\section{Results and Discussion}

\subsection{Determining strain at cavities}

\begin{figure*}[ht]
\centering
\includegraphics[width = \textwidth]{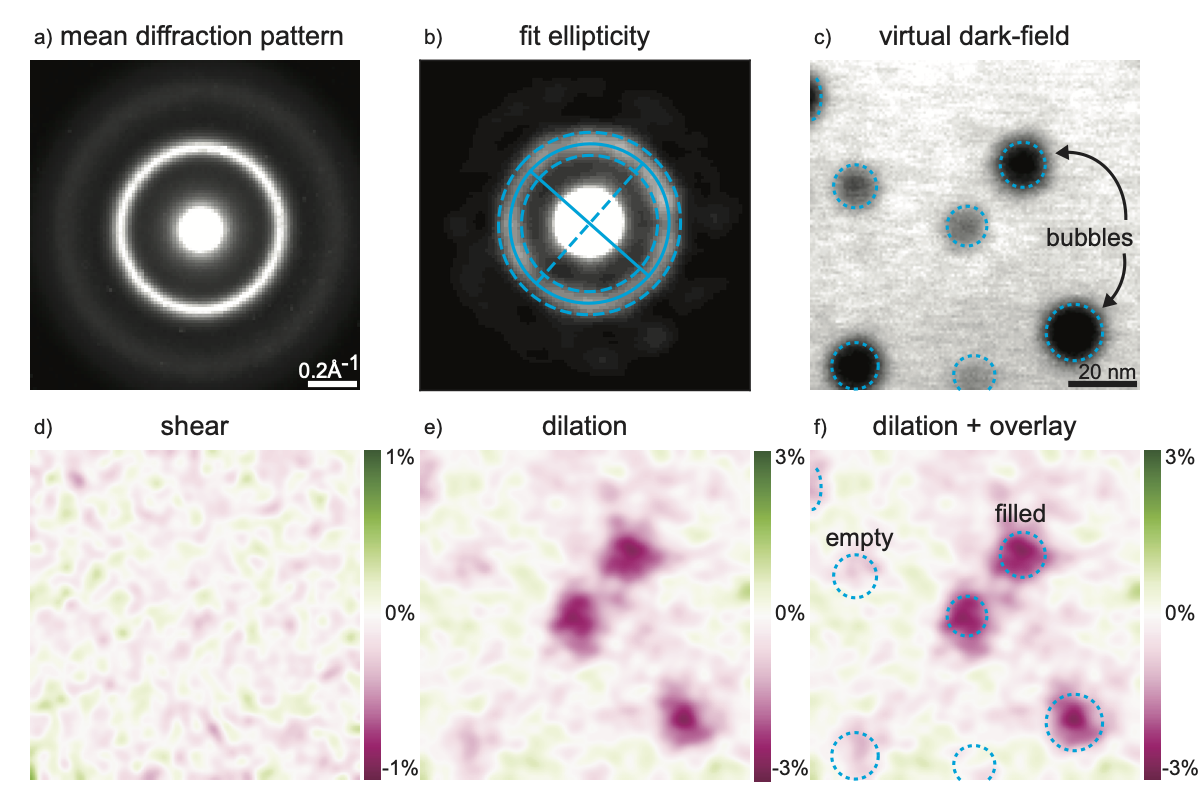}
  \caption{(a) Mean diffraction pattern. (b) The first amorphous ring of each diffraction pattern was fit with an ellipse. (c) Virtual dark-field shows location of all cavities. While there is little (d) shear strain, the (e) dilation maps show compression from filled bubbles. The (f) overlay highlights cavities that may be filled or unfilled.}
  \label{fig:strain}
\end{figure*}

In this study, a 4D-STEM configuration with a convergence angle of 0.25 mrad was employed to generate diffraction speckle patterns.
The data were collected with a probe step size of 0.75 nm, and the probe diameter was 1.4 nm, yielding a spatial resolution limited by the probe size.
This fine spatial resolution is critical for detecting subtle local variations in structure that would otherwise be averaged out in larger-scale measurements. In the case of an amorphous material, strain is simply the relative change in atomic distance between volumes of atoms~\cite{poulsen2005measuring,lunt2018origins}. Strain in amorphous materials induces local variations in atomic spacing, which disrupts the isotropy of scattering, causing distortions in the diffracted nanobeam patterns~\cite{voyles2002fluctuation}. These distortions appear as changes in the ellipticity of the diffracted ring, where the major and minor axes reflect the degree and directionally of local strain.

The diffraction patterns were recorded while rastering the beam across a sample of Ar-irradiated Bi$_2$O$_3$, which contains both Ar-induced voids (empty cavities) and bubbles (filled cavities).
The mean diffraction pattern (Figure~\ref{fig:strain}a), the average pattern from all probe positions, is shown in Figure~\ref{fig:strain}c, showing characteristic ring patterns of amorphous materials where the scattering vector reflects atomic spacings.
To create a position-sensitive strain map of this material, first a Gaussian blur with a kernel size of 3 pixels was applied to the diffraction patterns to remove the influence of high-frequency noise and small-scale speckle fluctuations.
The resulting smoothed diffraction patterns were analyzed using an elliptical ring fitting method, shown in Figure~\ref{fig:strain}b, allowing us to extract the local strain distribution on a probe-by-probe basis. 
The variations in the major and minor axes of the elliptical fits were subsequently used to calculate strain tensors, providing a detailed map of local strain across the Bi$_2$O$_3$ sample.
This approach is based on the tensorial form of the quadratic equation for an ellipse, as described in the Methods Section, which accurately describes the geometric distortions in the diffraction patterns caused by local strain variations.
We expect the precision of our strain calculation to be approximately 0.2 \% ~\cite{kennedy2020tilted}.

By calculating strain for each probe position in the diffraction pattern, we can predict which cavities are filled and which are empty.
The virtual dark-field from this area is shown in Figure~\ref{fig:strain}c, highlighting the morphology of the sample.
Comparing the strain maps to the dark-field image, there is very little shear strain (Figure~\ref{fig:strain}d) across the field of view, although there is significant dilation (Figure~\ref{fig:strain}e) in some cavities.
The overlay between the virtual dark-field image (Figure~\ref{fig:strain}f) and the strain map highlight that in this field of view, some cavities are filled bubbles. 
This technique can be compared to elemental mapping, shown in Figure~\ref{fig:overview}b, which also confirms a mix of filled and empty cavities, validating the results from these calculations. 
However, elemental mapping is a dose inefficient technique, especially compared to 4D-STEM, leading to sample damage during data acquisition. 
Our 4D-STEM analysis provides a new method to differentiate between voids and bubbles, while also capturing other structural information for further characterization without damaging the sample.

Although we can clearly observe some bubbles that are filled (Figure~\ref{fig:strain}f), for other cavities the presence or absence of Ar is less definitive. 
Since the beam interacts with all material in its path, roughly 40-50 nm, the strain measurements represent average values along the direction of propagation of the electron beam.
This means that, in areas with cavities, the strain measurements are influenced by both the material at the cavity interface and the surrounding amorphous Bi$_2$O$_3$. In the cavities that do not exhibit a large strain gradient, we suspect that these are cavities where the Ar gas has been released, possibly during the focused ion-beam (FIB) process as the material is bisected.
However, because the projection nature of this technique implies that strain represents signal from the entire cross section, it is possible that these are also small or less pressurized bubbles, where the signal is dominated by the surrounding unstrained matrix. 

In oxides, oxygen vacancy clusters are known to exhibit high mobility~\cite{andersson2012stability}, but the coarsening of gas bubbles indicates that the transport involves multiple species and is not limited to oxygen. Gas diffusion mechanism depend on the gas species and the size of the interstitial sites in the material. Noble gases are insoluble in solid materials, coalescing and binding within vacancies~\cite{han2024dissolution}. Bubble coarsening occurs through vacancy-mediated transport of gas and in the process, other species are displaced~\cite{yu1997defect,lin2023analysis}.
In the filled cavities (bubbles), the gas pushes against the surrounding material, decreasing the lattice parameter and introducing a compressive strain at the gas/matrix interface.
The majority of the strain is concentrated at the interface~\cite{xiao2020quantitative,wang2022modeling}.
However, for bubbles where the Ar gas has been released, the material relaxation and strain reduction is likely.
Although some ambiguity exists about voids with low strain gradients compared to the matrix, the method accurately measures projected strain variations across the region.

\subsection{Comparing local ordering between cavities and the matrix}

\begin{figure*}[ht]
\centering
\includegraphics[width = 0.95\textwidth]{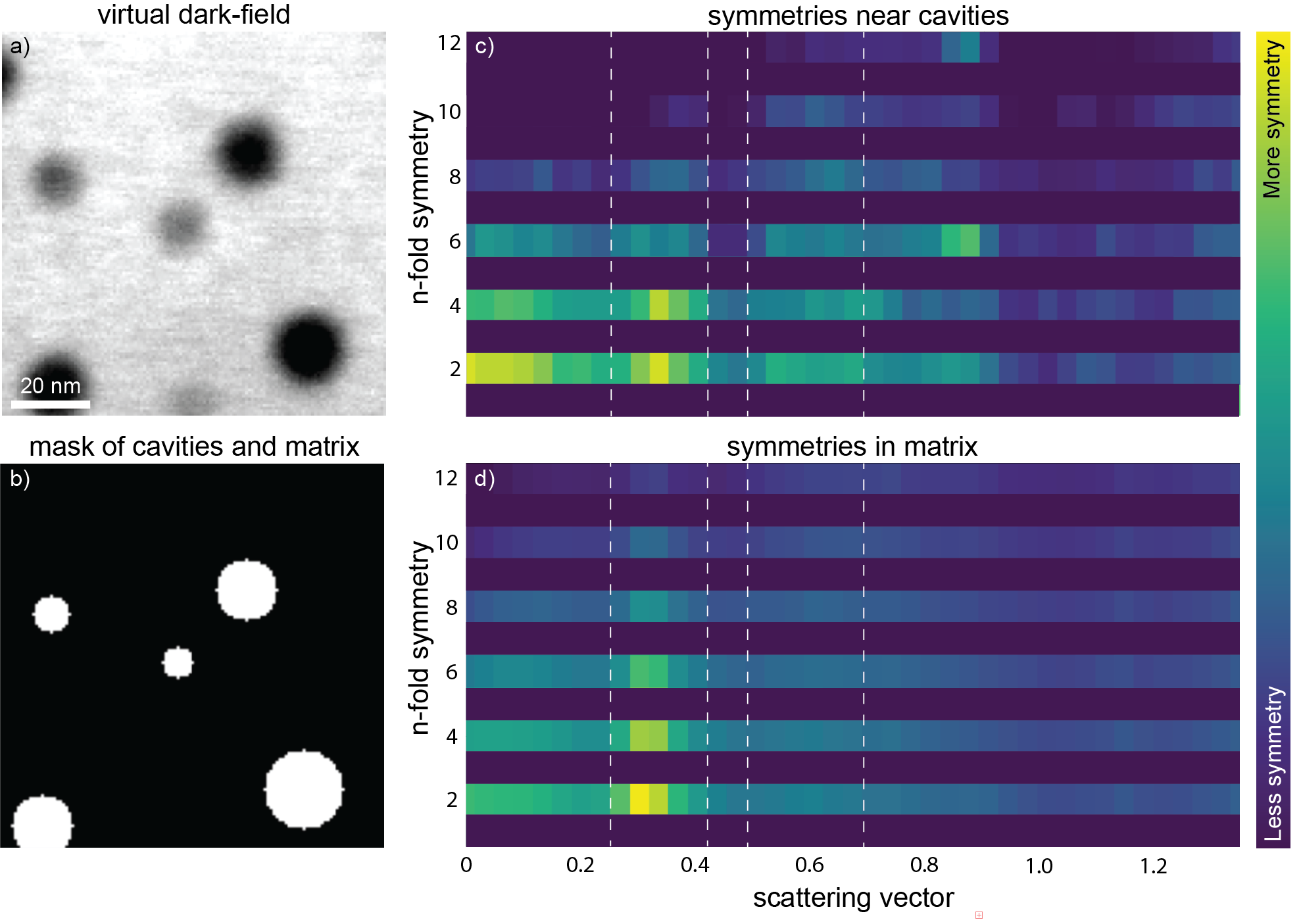}
  \caption{(a) A virtual dark-field image showing cavities in amorphous Bi$_2$O$_3$ matrix. (b) The mask used to create two classes, cavities and amorphous matrix, for symmetries analysis. (c) The even symmetries measured from the white masked cavities and the (d) black masked bulk material shown in (b). The white dashed lines indicate the first and second amorphous rings centered at 0.34 and 0.57 $\mathrm{\AA}^{-1}$, respectively. Higher intensities indicate greater amounts of specific \textit{n}-fold symmetries within the masked regions. Crystalline symmetries are most present within the first diffracted halo for both the matrix and the cavities, (d) however the cavities have more detected symmetries at higher scattering vectors, indicating a greater degree of paracrystalline ordering and greater structural heterogeneity at the cavity walls within the Bi$_2$O$_3$.}
  \label{fig:symmetry}
\end{figure*}

The versatility of 4D-STEM data means that we can analyze the symmetries as well as the strain from the same field of view.
To differentiate the cavities and matrix, a mask was created as shown in Figure~\ref{fig:symmetry}a,b.
Since the probe was parked at the center of the region in Figure~\ref{fig:symmetry}a (marked in Fig. S1 of the Supplemental Information), the diffraction patterns collected from this area were excluded from the symmetry analysis to prevent potential beam-induced recrystallization or structural changes from influencing the analysis of either the matrix or the cavity regions.
Symmetry calculations followed the procedure outlined by Liu \textit{et al.} (2015)\cite{liu2015interpretation}, where the angular cross-correlation function (CCF)~\cite{wochner2009xray,bojesen2020statistical} was employed to detect subtle crystalline symmetries embedded in the diffraction patterns on a pattern-by-pattern basis.
As noted in previous studies, odd symmetries can appear erroneously in this type of analysis due to dynamical scattering, and to a lesser extent, from inherent aberrations in the electron beam.
While methods exist to minimize these effects, the Bi$_2$O$_3$ lamellae used in this study are significantly thicker than those typically analyzed using this approach, making multiple scattering events contributing to dynamical diffraction more prominent.
Therefore, odd symmetries were excluded from the analysis (Figure~\ref{fig:symmetry}c,d). Further details of the methodology can be found in Liu \textit{et al.} (2015).

Figure~\ref{fig:symmetry}c,d shows the results of the \textit{n}-fold symmetry analysis for the cavity region and the amorphous matrix, respectively.
The dashed lines indicate the locations of the first and second amorphous rings, centered at 0.34 and 0.57 $\mathrm{\AA}^{-1}$, respectively.
The most significant differences between the two regions are that symmetries are almost exclusively detected at the first amorphous ring for the matrix, while they are observed at additional scattering vectors for the cavities.

For the amorphous matrix, two-fold symmetry is dominant in the amorphous halo, with a decrease in the intensity of \textit{n}-fold symmetries as \textit{n} increases, particularly at the first ring scattering vector.
This trend is expected, as two-fold symmetry is the most commonly observed symmetry in amorphous materials due to the inherent lack of long-range order in these structures~\cite{liu2013systematic,huang2022correlation}. 
Additionally, two-fold symmetry tends to be the most prevalent in the cavity regions.
However, an interesting deviation from this pattern occurs at lower scattering vectors (less than 0.1 $\mathrm{\AA}^{-1}$), where two-fold symmetries are still prominent, and at higher scattering vectors beyond the first amorphous ring, suggesting a more complex structural environment within the cavities.
Notably, at a scattering vector near 0.9 $\mathrm{\AA}^{-1}$, there is clear evidence of six-fold and 12-fold symmetries, along with a lesser but still detectable presence of two-fold symmetry.
These higher-order symmetries, which are often associated with crystalline structures, indicate the presence of paracrystalline ordering within the amorphous material.
Paracrystallinity refers to SRO that resembles a crystalline structure but with random positional deviations, often giving rise to diffraction features similar to those of amorphous materials with enhanced local symmetry~\cite{voyles2001structure}.

The observation of such crystalline-like symmetries within what is nominally an amorphous structure suggests a higher degree of ordering within the cavity regions compared to the matrix.
This heightened degree of ordering can be interpreted as evidence of a more structured local arrangement in the cavities, potentially due to the confined nature of the cavity environment promoting local paracrystalline ordering.
The increased prevalence of even symmetries at the cavity interfaces, particularly at higher scattering vectors, points to the formation of inchoate crystallites — small, partially ordered regions of the material.
This is in contrast to the more disordered matrix, where the lack of detectable higher $\textit{n}$-fold symmetries underscores the overall amorphous character of the material.

Taken together, the observations that cavity shells exhibit higher compressive strain and that more ordered crystalline symmetries are present within the Bi$_2$O$_3$ at the cavity interfaces suggest that gas bubbles in amorphous materials can induce significant localized structural modifications.
The mobility of Bi and O atoms near the bubbles leads to localized crystallinity enhancements and denser atomic packing.
Understanding this localized strain can also provide insights into how amorphous materials might evolve or degrade under stress or irradiation.
During irradiation, the energetic transfer from Ar ions to Bi and O atoms is sufficient to disrupt the existing crystal structure, leading to amorphization.
However, as bubbles coarsen, the enhanced ion mobility facilitates the rearrangement of Bi and O atoms into more thermodynamically favorable, partially ordered configurations.
Although this reordering is incomplete at the cavity edges, it results in local ordering that is distinctly more ordered than the surrounding amorphous matrix, suggesting a dynamic interplay between bubble formation, strain, and atomic rearrangement in the material.
In short, this complex structure arises from the interplay between gas pockets and the surrounding material influences, which can impact the material's properties, such as its mechanical strength, diffusion behavior, or conductivity. 

\subsection{Estimating pressure inside filled cavities}

In irradiated materials, the pressure of the gas within a bubble often balances the osmotic pressure at the surface. In such cases, the pressure $\textit{p}$ is related to the surface energy $\gamma$ and the bubble radius $\textit{a}$ by the equation $\textit{p}$ = 2$\gamma$/$\textit{a}$. This relationship assumes an equilibrium condition between the bubble and the surrounding material~\cite{donnelly1985density}.

For the bubbles in the region selected for strain analysis, the radii range from 5.6 to 8.2 nm.
Using a nominal radius of 6 nm and a surface energy of 0.3 J/m$^2$ for Bi$_2$O$_3$ (within the reported range of 0.2 to 0.8 J/m$^2$)~\cite{lei2013density,guenther2014size,fujino2005surface}, the estimated internal pressure in the bubble is approximately 0.1 GPa.
However, this calculation assumes equilibrium, which does not apply in this case.
As shown in the strain maps in Figure~\ref{fig:strain}, the material is not in equilibrium.
The compression observed at the cavity edges indicates that the pressure inside the bubbles exceeds the equilibrium value. Therefore, a more accurate estimate of the internal pressure can be derived from the measured strain fields, though it remains an approximation.

From linear elasticity theory we developed an approximate lower bound (see Methods Section for details) for the internal pressure of a representative 6 nm radius bubble with a maximum dilatational strain of approximately $\langle \epsilon_d\rangle=$-3\% at the surface of the bubble:
\begin{equation}
    p \gtrsim \frac{2\gamma}{a}-4\mu \langle \epsilon_d \rangle,
\end{equation}

This gives an internal pressure of 2.5 GPa. This estimate appears reasonable, as the internal pressure of irradiation gas-induced bubbles, which depends on bubble size and other factors, typically ranges from a fraction of 1 GPa to several tens of GPa ~\cite{donnelly1985density,dowek2021determination}.
It is important to note that this is a rough estimate, intended to demonstrate the reasonableness of the strain measurements and their utility in estimating the internal gas pressure within bubbles in solid materials.

\section{Methods}

\subsection{Sample preparation}

99.9\% purity Bi$_2$O$_3$ sputtering targets were purchased from the Kurt J. Lesker Company and prepared for irradiation as described in Kennedy et al. (2024)~\cite{kennedy2024insights}. Ion irradiations were performed using a 200 kV Danfysik Ion Implanter in the Ion Beam Materials Laboratory (IBML) at Los Alamos National Laboratory. The samples were mounted on a Ni-block with double-sided carbon tape for good thermal contact and active cooling ensured that the target temperature during the irradiation remained below 35$^{\circ}$C. The target chamber vacuum was maintained near 1 $\times$ 10$^{-7}$ torr. The irradiation was performed using Ar$^+$ ions at 400 keV with a fluence of 1 $\times$ 10$^{16}$ Ar ions/cm$^2$. SRIM damage profile for the region in Figure~\ref{fig:overview}a is provided in Fig. S2 of the Supplemental Information and the methods for determining the damage profile are outlined in Kennedy et al. (2024)~\cite{kennedy2024insights}.

Bi$_2$O$_3$ cross-sections for TEM analysis were prepared using a Helios NanoLab 600 dual-beam SEM-FIB. Thinning was performed iteratively on both sides of the lamellae with decreasing acceleration voltages and currents. Final thinning was performed at 2 keV with a current of 86 pA to minimize material redeposition and damage, resulting in a thickness range of approximately 30 nm at the surface to 80 nm at the amorphous/crystalline interface~\cite{kennedy2024insights}.

\subsection{Bright field TEM and EDS}

TEM images and energy-dispersive x-ray spectroscopy (EDS) maps, Figure~\ref{fig:overview} were acquired on an image-corrected FEI Titan operated at 300 keV with a OneView 4K CCD camera. Bright field images are intentionally underfocused and a 100 $\mu$m objective aperture was used to increase feature contrast, specifically of the Ar bubbles and voids in the amorphous matrix. Previously reported grazing incident X-ray diffraction (GIXRD) collected from the sample sample supports that Ar$^+$ irradiation at 400 keV results in an amorphous layer on the sample surface~\cite{kennedy2024insights}. Additional EDS maps are are provided in Fig. S3 of the Supplemental Information.

\subsection{4D-STEM data acquisition}

4D-STEM was acquired on the TEAM I microscope at the Molecular Foundry, a modified FEI Titan double-
aberration-corrected microscope.
The TEAM I was operated at 300 kV with a probe defined by 5 $\mu$m aperture, making a  0.25 mrad convergence angle.
Data was acquired with approximately 2.3$\times 10^{3}$ e$^-$/\AA $^2$ on the Dectris Arina camera operating in full-frame (196$\times$196) mode.

\subsection{Strain analysis}

To measure strain, the amorphous halo of each diffraction pattern in the 4D-STEM dataset was fit with a two-sided Gaussian model plus a background Gaussian function in the open-source package \texttt{py4DSTEM}~\cite{savitzky2021py4dstem}. 
Before fitting each diffraction pattern was Gaussian filtered.
The fit produced the elliptical parameters for each halo, namely a major axis ($\alpha$), a minor axis ($\beta$), and an angular offset ($\theta$). 
These parameters were converted to $A$, $B$, and $C$ defining an equation of an ellipse: 
\begin{equation}
Ax^2 + Bxy + Cy^2 = 1.
\end{equation}

$A$, $B$, and $C$ were calculated as follows: 
\begin{align}
    A &= \alpha^2 \sin^2(\theta) + \beta^2 \cos^2(\theta) \\
    B &= 2(\alpha^2 - \beta^2) \sin(\theta)\cos(\theta) \\ 
    C &= \alpha^2 \cos^2(\theta) + \beta^2 \sin^2(\theta).
\end{align}

The eigendecomposition of the matrix of the quadratic form
\begin{align}
 \bm{m}_{\text{ellipse}} =\begin{bmatrix}
A & B/2 \\
B/2 & C
\end{bmatrix}
\end{align}

is 
\begin{align}
\bm{m}_{\text{ellipse}} = \bm{V} \bm{\Lambda} \bm{V}^{-1}.
\end{align}
 The angle $\phi$ between the principal axis and the reference frame is defined as
\begin{align}
\phi = \tan^{-1} \left( \frac{\bm{V}_{1,0}}{\bm{V}_{0,0}} \right),
\end{align}

creating the rotation matrix: 
\begin{align}
    \bm{m}_{\text{rot}} = \begin{bmatrix}
\cos(\phi) & -\sin(\phi) \\
\sin(\phi) & \cos(\phi)
\end{bmatrix}.
\end{align}

Given that the square root of the eigenvalues are the lengths of the major and minor axes, the transformation matrix is then: 
\begin{align}
\bm{m}_{\text{trans}} = \bm{m}_{\text{rot}} ~ \bm{\Lambda} ^{1/2} ~ \bm{m}_{\text{rot}}^T.
\end{align}

Finally a strain matrix ($\bm{m}_{\text{strain}}$) was produced by multiplying the transformation matrix by the inverse of a transformation matrix calculated in the same way but from a reference region ($\bm{m}_{\text{ref}}$). The strain matrix is: 
\begin{align}
\bm{m}_{\text{strain}} = \bm{m}_{\text{trans}} \bm{m}_{\text{ref}}^{-1},
\end{align}

where we define

\begin{align}
\bm{m}_{\text{strain}} = \begin{bmatrix}
\epsilon_{xx} + 1 & \epsilon_{xy'} \\
\epsilon_{xy''} & \epsilon_{yy} + 1
\end{bmatrix}
,
\end{align}
such that

\begin{align}
\epsilon_{xy} = \frac{1}{2} (\epsilon_{xy'} + \epsilon_{xy''}).
\end{align}

In  Fig.~\ref{fig:strain}, we plot dilation as $\epsilon_{xx} + \epsilon_{yy}$ and shear as $\epsilon_{xy}$.

\subsection{Probing crystalline symmetries}

Despite lacking long-range order, amorphous materials exhibit localized atomic ordering at medium- and short-range length scale~\cite{treacy2005fluctuation}.
In contrast, small volume diffraction measurements preserve these local variations and enable the study of higher-order atomic correlations.
Two techniques — fluctuation electron microscopy and electron nanodiffraction — are used to access these higher-order correlations of local atomic clusters~\cite{liu2013systematic,treacy2005fluctuation}. However, dynamical diffraction complicates the interpretation of nanodiffraction patterns from disordered materials, especially in the presence of lens aberrations or multiple scattering.

As described in Liu et al. (2015)~\cite{liu2015interpretation}, the angular symmetries in the electron nanodiffraction patterns of amorphous Bi$_2$O$_3$ are measured.
Dynamical diffraction and lens aberrations distort the diffracted volume and complicate symmetry interpretation.
To quantify symmetries, the angular cross-correlation function (CCF) of the nanodiffraction patterns is analyzed~\cite{wochner2009xray}~\cite{gibson2010substantial}. The four-point CCF is defined as a function of probe position \textit{\textbf{r} = (x,y)} and scattering vector \textit{k}

\begin{align}
C(\textbf{r},k,\Delta) = \frac{\langle I(k,\varphi)I(k,\varphi+\Delta) \rangle - \langle I(k,\varphi)\rangle^2}{\langle I(k,\varphi)\rangle^2}
\end{align}

where \textit{I(k,$\varphi$)} is the diffracted intensity at a specific scattering vector that is averaged over the azimuthal angle $\varphi$.

Decomposing the CCF into a Fourier cosine series, symmetry magnitudes as a function of the scattering vector are extracted.
Analysis is limited to even symmetries as odd symmetries are mainly the result of kinematic diffraction.
The diffraction patterns change depending on whether the probe matches the size of the local cluster~\cite{liu2013systematic}.
By spatially averaging the symmetry magnitudes across the bulk amorphous and cavity portions of the sample, the dominant local structures are mapped for each region.
These symmetry maps provide insight into the spatial distribution of SRO in the material.

\subsection{Estimating bubble pressure}

Measured strain values enable us to estimate the pressure within a bubble, and help us define an equation of state for the bubble. The accuracy of this method is dependent on the resolution of the data and the accuracy of the shear modulus value used in the calculation. 

The elastic displacement field generated by a bubble of radius $\textit{a}$ and pressure $\textit{p}$ in a solid isotropic elastic medium with shear modulus $\mu$ is given in spherical coordinates by~\cite{wolfer1988pressure,Chalon2004}

\begin{equation}
    u_{r} = \frac{a^3\Delta p}{4\mu r^2},
\end{equation}
where $\mu$ is the shear modulus, $\textit{r}$ is the radial distance from the center of the bubble and the other components, associated with the polar and azimuthal angles, are zero. $\Delta p$ is the pressure experienced by the material at the surface of the bubble, and can be shown to $\Delta p = p-2\gamma/a$~\cite{WILLIS1969solid}.
As the linear elastic strain is defined as $\bm{\epsilon} = \frac{1}{2}\left(\nabla\bm{u}+(\nabla\bm{u})^T\right)$, we can express the dilational strain on the xy plane at a point exterior to a single bubble  as
\begin{equation}
    \epsilon_d=\epsilon_{xx}+\epsilon_{yy} = \frac{a^3\Delta p}{4\mu}\left(\frac{2r^2-3(x^2+y^2)}{r^5}\right),
\end{equation}

where $r=\sqrt{x^2+y^2+z^2}$. The strain fields shown in Figure \ref{fig:strain} represent a weighted average of the strains through the thickness of the material. If the material is of thickness $2l$ and we assume that the bubble is at $z=0$ and at distances $l(1-\chi)$ and $l(1+\chi)$ from the two sample surfaces we can express the weighted sum as

\begin{equation}\label{eq:integrate}
    \langle\epsilon_d\rangle(x,y,\chi) = \frac{1}{2l}\int_{-l(1-\chi)}^{l(1+\chi)}\epsilon_d(x,y,z)w(z)dz,
\end{equation}
where $w(z)$ is a weighting function. As the exact weighting is unknown we aim to develop approximate bounds on strain. We consider the two extremes where $w^{min}(z)=1$ and $w^{max}(z)=2l\delta(z)$, with $\delta(z)$ being the Kronecker-Delta function. For further simplicity, let us consider strain at points on the equator of the bubble, $\sqrt{x^2+y^2}=a$. Substitution of $w^{max}$ into Eq. \eqref{eq:integrate} gives a constant value while in the case of $w^{min}$, $\langle\epsilon_d^{min}\rangle = \langle\epsilon_d^{min}\rangle(\chi)$, with a maximum value at $\chi=0$. Taking $\chi=0$ we find:

\begin{subequations}\label{eq:bounds}
    \begin{align}
        \langle \epsilon_d^{max}\rangle &= -\frac{\Delta p}{4\mu},\\
        \langle \epsilon_d^{min}\rangle &= -\frac{\Delta p}{4\mu}\frac{a^3}{\left(a^2+l^2\right)^{3/2}}.\label{eq:upper-bound}
    \end{align}
\end{subequations}

\begin{figure}[ht]
\centering
\includegraphics[width = 0.42\textwidth]{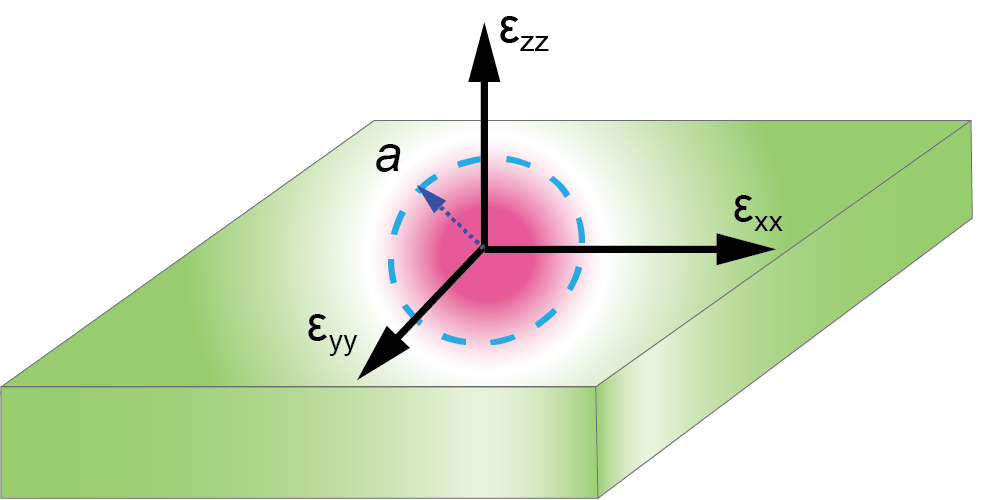}
  \caption{Schematic representation of the strain field surrounding an Ar bubble projected in the \textit{xy}-plane where \textit{a} is the radius of the bubble.}
  \label{fig:pressure}
\end{figure}

The shear modulus of amorphous Bi$_2$O$_3$ is not documented; however, the shear modulus of $\alpha$-Bi$_2$O$_3$ is reported in the 20-30 GPa range~\cite{gomis2019elastic,jain2013commentary}. The amorphous phase of Bi$_2$O$_3$ is expected to have a lower shear modulus compared to the crystalline phase~\cite{knuyt1991calculation}. To approximate the pressure in the bubbles, a shear modulus, $\mu$, of 20 GPa is used as an estimate. A nominal bubble radius of 6 nm is used to exemplify the internal pressure of the bubble. If we assume a sample thickness of $2l=50$ nm and $\langle\epsilon_d\rangle = -3\%$, Eq. \eqref{eq:upper-bound} gives $\Delta p = 189$ GPa. This is an extremely high value for bubble pressure. As the bubble approaches the surface for the same measured value of $\langle\epsilon_d\rangle$, $\Delta p$ decreases dramatically. However, as the bubble approaches the surface interactions between the bubble and surface, which are not considered here, become more important~\cite{WILLIS1969solid,WILLIS1975interaction}, leaving the development of a more effective upper bound on bubble pressure as future work.

\section{Conclusions}

Using 4D-STEM data, this work demonstrates that gas bubbles in amorphous Bi$_2$O$_3$ result in compressive strain and induce paracrystalline ordering within the local atomic structure. Atomic spacing in the amorphous matrix is measured to dilate by roughly 3\% at the edge of Ar bubbles. The analysis of strain fields provides a means of estimating internal gas pressures, revealing that these pressures are significantly higher than the equilibrium value, with a lower bound estimate of 2.5 GPa. These findings underscore the importance of considering localized structural heterogeneity in amorphous materials, particularly under stress or irradiation. 

The ability to capture subtle structural features, such as the emergence of crystalline symmetries and strain variations, offers new insights into the behavior of disordered materials and their response to external perturbations. The correlation between bubble formation, strain, and induced ordering suggests that amorphous materials can exhibit structural memory effects, which may influence their mechanical integrity and transport properties. Tracking strain evolution at the nanoscale provides a pathway for predicting failure mechanisms and critical strain thresholds, which is particularly relevant for cladding materials in nuclear reactors, where irradiation-induced swelling and embrittlement can lead to mechanical degradation. The pressure-strain relationships observed here parallel effects seen in other irradiation-driven transformations, such as metal transmutation~\cite{lloyd2024microstructural} and hydride formation~\cite{jang2017effect}, where internal stresses play a key role in phase stability and mechanical failure. Furthermore, these findings highlight the role of nanoscale heterogeneity in governing macroscopic properties such as ionic and electronic transport, making them particularly relevant for applications in radiation-tolerant materials, solid-state electrolytes, and functional oxides. The methods presented here provide a versatile approach for studying the complex interplay between structural defects, gas inclusions, and material properties, with implications for both the degradation and potential functionalization of amorphous materials. Ultimately, this study contributes to a more nuanced understanding of the behavior of amorphous materials and offers a framework for future research into their stability and performance in extreme environments.

\section{Acknowledgments}

This work was primarily supported by the Laboratory Directed Research and Development program of Los Alamos National Laboratory under project number 20240879PRD4. This work was supported by the U.S. Department of Energy, Office of Science, Basic Energy Sciences, Materials Sciences and Engineering Division. Los Alamos National Laboratory is operated by Triad National Security, LLC, for the National Nuclear Security Administration of U.S. Department of Energy (Contract No. 89233218CNA000001). This work was performed, in part, at the Center for Integrated Nanotechnologies, an Office of Science User Facility operated for the U.S. Department of Energy (DOE) Office of Science. The Los Alamos National Laboratory, an affirmative action equal opportunity employer, is managed by Triad National Security, LLC for the U.S. Department of Energy's NNSA, under contract 89233218CNA000001. Sandia National Laboratories is a multimission laboratory managed and operated by National Technology \& Engineering Solutions of Sandia, LLC, a wholly owned subsidiary of Honeywell International Inc., for the U.S. Department of Energy’s National Nuclear Security Administration under contract DE-NA0003525. This paper describes objective technical results and analysis. Any subjective views or opinions that might be expressed in the paper do not necessarily represent the views of the U.S. Department of Energy or the United States Government. Work at the Molecular Foundry was supported by the Office of Science, Office of Basic Energy Sciences, of the U.S. Department of Energy under Contract No. DE-AC02-05CH11231. S.M.R. and C.O. acknowledge support from the U.S. Department of Energy Early Career Research Award program. The authors thank Chengyu Song at the Molecular Foundry for maintenance of and assistance with the TEAM I TEM and James Valdez at LANL for sample polishing.

\section{Data Availability}
Data will be made available on request.

\bibliography{main}

\end{document}


\maketitle

\pagestyle{empty}

\begin{figure*}[!ht]
\centering
\includegraphics[width=0.52\textwidth]{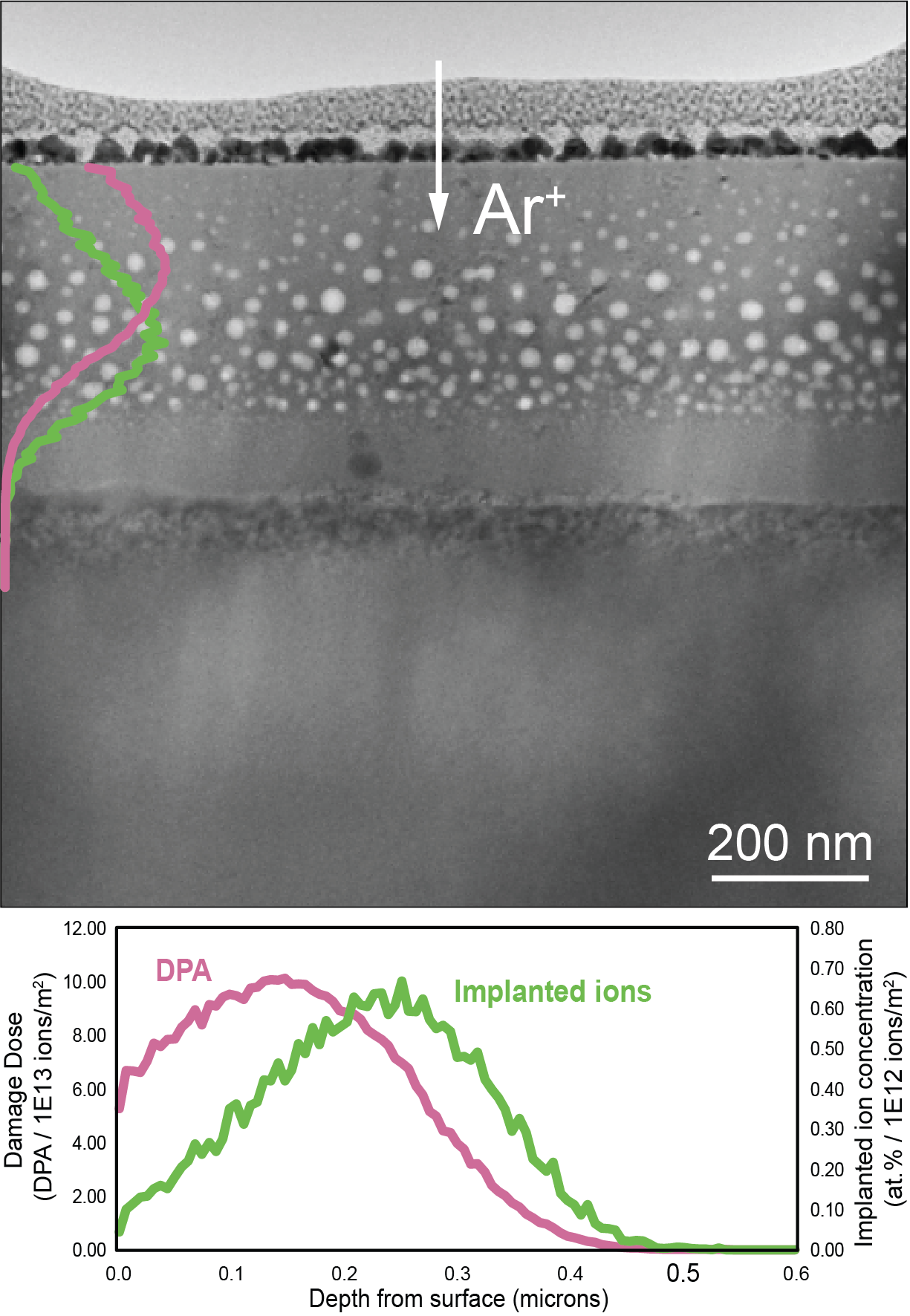}
\caption{\label{fig:figs1} Bright-field TEM image of cross-section after Ar irradiation replicated from Fig. 1a with overlain dsiplacements per atom (DPA) and implanted ion concentration profiles. The damage profiles, calculated in SRIM, show peak implantation and DPA within the amorphous layer. Beneath the amorphous region, there is a layer of dislocation loops and, beneath that, the pristine crystal.}
\end{figure*}

\begin{figure*}[!ht]
\centering
\includegraphics[width=0.70\textwidth]{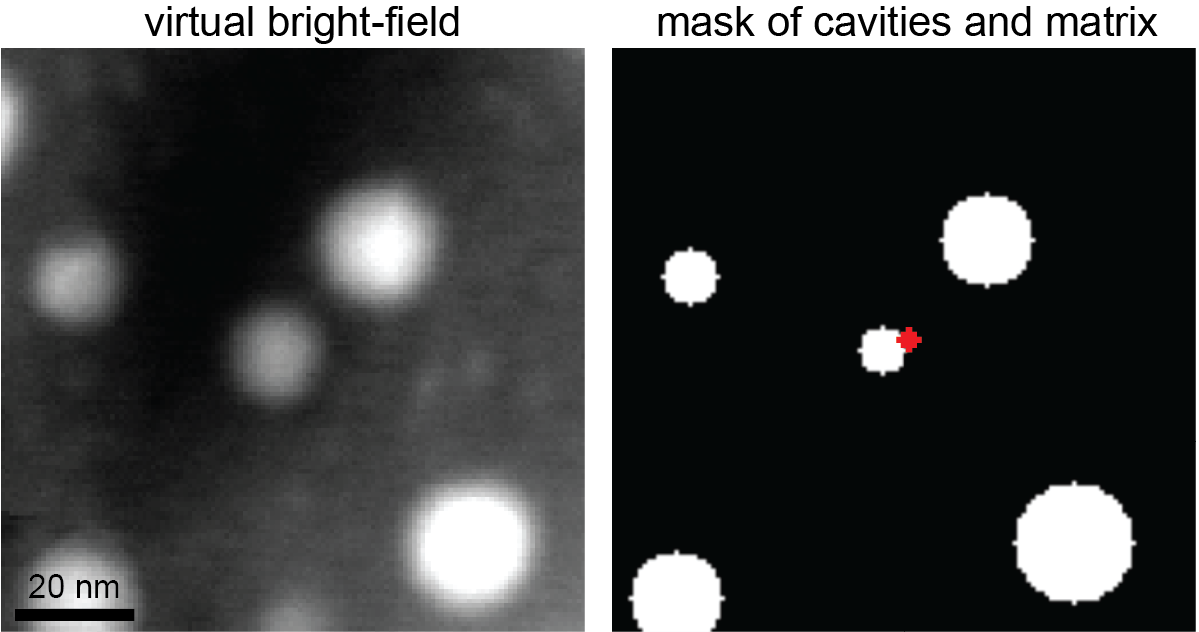}
\caption{\label{fig:figs1} Virtual bright-field image of region analyzed in Figures 2 and 3 and the mask shown in Figure 3b replicated. Pixels belonging to the bulk amorphous matrix and the cavity interfaces are separated using the mask. The red marking in the center of the mask region indicates where the electron beam was parked momentarily at the end of scans. These pixels were excluded from both the cavity and bulk classes due to the possibility of structural transformation from material interactions with the beam.}
\end{figure*}

\begin{figure*}[!ht]
\centering
\includegraphics[width=0.98\textwidth]{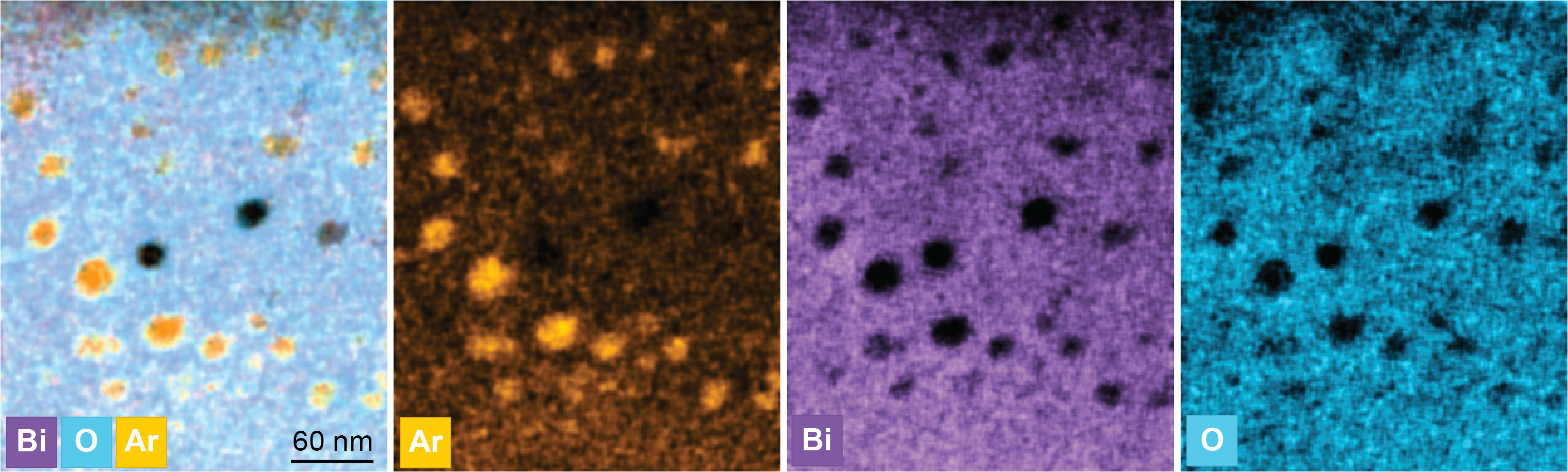}
\caption{\label{fig:figs1} Composite and separate EDS maps for the irradiating species, Ar, and the components of the amorphous matrix, Bi and O.}
\end{figure*}
